\documentclass[a4paper,12pt]{article}

\usepackage{graphics, epsfig}

\begin{document}

\author{C. Barrab\`es\thanks{E-mail : barrabes@celfi.phys.univ-tours.fr}\\
\small Laboratoire de Math\'ematiques et Physique Th\'eorique,\\
\small CNRS/UMR 6083, Universit\'e F. Rabelais, 37200 TOURS,
France\\P. A. Hogan\thanks{E-mail : phogan@ollamh.ucd.ie}\\ \small
Mathematical Physics Department,\\ \small  National University of
Ireland Dublin, Belfield, Dublin 4, Ireland}

\title{Deflection of Highly Relativistic Particles in a Gravitational Field}
\date{}
\maketitle

\begin{abstract}
A novel approach to the calculation of the deflection of highly
relativistic test particles in gravitational fields is described.
We make use of the light-like boosts of the gravitational fields
of the sources. Examples are given of the deflection of highly
relativistic particles in the Schwarzschild and Kerr gravitational
fields, in the field of a static, axially symmetric, multipole
source and in the field of a cosmic string. The deflection of
spinning particles is also discussed.

\end{abstract}
\vskip 2truepc\noindent PACS number(s): 04.20.Jb
\thispagestyle{empty}
\newpage
\section{Introduction}\indent
In a recent paper \cite{BH} we took a fresh look at the
light--like boost of the Schwarzschild gravitational field which
was originally worked out in an important paper by Aichelburg and
Sexl \cite{AS}. Our point of view entailed emphasising the role of
the Riemann tensor rather than the metric. We illustrated this by
calculating in \cite{BH} the light-- like boost of a Weyl static,
axially symmetric gravitational field. More recently we have
described the light--like boost, from our point of view, of domain
walls and cosmic strings \cite{BHI} and of the Kerr gravitational
field \cite{BH2}. We present here a novel application of some of
these ``boosted gravitational fields'' by using them to study the
deflection of highly relativistic particles in the original
``unboosted'' gravitational fields. This in turn can shed some
light on properties of the boosted gravitational fields
themselves. The particles we consider are ultrarelativistic in the
sense that they are travelling with speed close to the speed of
light. Hence our point of view involves an approximation in which
the leading terms in the angles of deflection that we calculate
are the most accurate for large impact parameter.

To illustrate and test our point of view we apply it, in section
2, to the simple case of the scattering of a high speed charge off
a nucleus in electromagnetic theory. We obtain the well known
angle of deflection in this case. In section 3 we give a general
description of the deflection of a high speed test particle in
gravitational fields modeled by a class of spacetimes which
include the light--like boosted Schwarzschild, Kerr and Weyl
space--times and the space--time of a boosted cosmic string. This
is used in section 4 to provide four examples: deflection of a
high speed test particle in the Schwarzschild field, the Kerr
field, the Weyl gravitational field of an isolated, static,
axially symmetric multipole source, and the field of a cosmic
string. The leading terms in the angles of deflection that we
obtain coincide with Einstein's angle of deflection of a photon
path in the Schwarzschild field and with the angle of deflection
obtained, for example, by Boyer and Lindquist \cite{BL} for a
photon equatorial path in the Kerr field. In the case of a Weyl
field we demonstrate that there are two contributions to the
deflection. There is a deflection \emph{out} of a plane orthogonal
to the axis of symmetry due to the presence of the \emph{odd}
multipole moments of the source and a deflection \emph{within}
such a plane due to the \emph{even} multipole moments, as one
would expect. In section 5 we discuss the deflection of a high
speed particle having intrinsic spin. A solution is given for the
case of a spinning particle (moving in the equatorial plane of the
source) when the spin is aligned with the symmetry axis.

\setcounter{equation}{0}
\section{Electromagnetic Example}\indent
We wish to consider the deflection of a high
speed charge $e'$ off a nucleus carrying a charge $e$. In the rest
frame $\bar{\cal S}$ of the nucleus we place $e$ at the spatial
origin and consider $e'$ projected with 3--velocity $v$ in the
positive $\bar x$--direction.
We shall use units in which the velocity of light $c=1$. Hence for
a high speed encounter we consider $v$ to be close to unity. In
the rest frame ${\cal S}$ of the charge $e'$ the nucleus appears
to be travelling in the negative $x$--direction with speed close
to the speed of light. We will consider what happens if,
with respect to the frame ${\cal S}$, the nucleus of charge $e$
is boosted to the speed of light. In this case the
observer in the frame ${\cal S}$, using the rectangular Cartesian
coordinates and time $x^\mu =(x, y, z, t)$, sees an
electromagnetic field given via the 2--form \cite{BH}
\begin{equation}\label{2.1} F=\frac{2\,e}{\rho
^2}\,\delta (x+t)\,(y\,dy+z\,dz)\wedge (dt+dx)\ ,
\end{equation}
with $\rho =\sqrt{y^2+z^2}$ and $\delta (x+t)$ is the Dirac delta
function which is singular on the null hyperplane $x+t=0$. This is a plane
impulsive electromagnetic wave travelling in the negative
$x$--direction having the null hyperplane $x+t=0$ as its history
in Minkowskian space--time with line--element
\begin{equation}\label{2.2}
ds^2=dx^2+dy^2+dz^2-dt^2\ .
\end{equation}
Bergmann \cite{B} was the first to observe that to a high
speed observer the Coulomb field of a point charge
increasingly resembles the field of an impulsive
electromagnetic wave as the speed of the observer relative to the charge
approaches the speed of light.
We see from (\ref{2.1}) that in addition to the delta function singularity
the field is also singular at $\rho
=0$ on $x+t=0$, which is one of the null geodesic generators of
the null hyperplane.

The Lorentz equations of motion for the
time--like world--line $x^\mu =x^\mu (s)$ of a charge $e'$ with
rest--mass $m$ encountering the electromagnetic wave (\ref{2.1})
are given by
\begin{eqnarray}\label{2.3}
\ddot x&=&-\frac{2e\,e'}{m\rho ^2}\,(y\,\dot y+z\,\dot
z)\,\delta (x+t)\ ,\nonumber\\ \ddot y&=&\frac{2e\,e'}{m\rho ^2
}\,y\,(\dot x+\dot t)\,\delta (x+t)\ ,\nonumber\\ \ddot
z&=&\frac{2e\,e'}{m\rho ^2}\,z\,(\dot x+\dot t)\,\delta (x+t)\
,\nonumber\\ \ddot t&=&\frac{2e\,e'}{m\rho ^2}\,(y\,\dot
y+z\,\dot z)\,\delta (x+t)\ ,\nonumber\\
\end{eqnarray}
with the dots indicating differentiation with respect to
proper--time $s$ along the world--line of the charge $e'$.
As a consequence of (\ref{2.3}) we have the first integral
\begin{equation}\label{2.4}
\dot x^2+\dot y^2+\dot z^2-\dot t^2=-1\ .
\end{equation}
Adding the first and last equations in (\ref{2.3}) gives $\ddot
x+\ddot t=0$ and so
\begin{equation}\label{2.5}
x+t=C\,s\ ,\end{equation} where $C$ is a constant of integration.
We have chosen to have $s=0$ on the world--line of $e'$ when it
intersects the null hyperplane $x+t=0$. We take $s<0$ on the
world--line of $e'$ to the past of $x+t=0$, and $s>0$ to the
future of $x+t=0$. Using (\ref{2.5}) we can write $\delta
(x+t)\,=\,C^{-1}\, \delta (s)$ and this can be substituted into
(\ref{2.3}). For $s<0$ and $s>0$ we see from (\ref{2.3}) that the
world--line of the charge $e'$ is a time--like geodesic of the
Minkowskian space--time with line--element (\ref{2.2}). Hence, and
in view of the right hand sides of (\ref{2.3}), we look for
solutions of (\ref{2.3}) and (\ref{2.4}) of the form
\begin{eqnarray}\label{2.6}
x&=& x_0 \,+\, x_1 s\,+\, X_1 s\, \theta (s)\, ,\nonumber\\
y&=& y_0 \,+\, y_1 s\,+\, Y_1 s\, \theta (s)\, ,\nonumber\\
z&=& z_0\, +\, z_1 s\, +\, Z_1 s\, \theta (s)\, ,\nonumber\\
t&=& -x_0 \,+\, t_1 s\,-\, X_1 s\, \theta (s)\, .\nonumber\\
\end{eqnarray}
Here $x_0$, $x_1$, $X_1$, $y_0$, $y_1$, $Y_1$, $z_0$, $z_1$, $Z_1$
and $t_1$ are constants and $\theta (s)$ is the Heaviside step function
which is equal to unity if $s>0$ and equal to zero if $s<0$.
We note from (\ref{2.6}) that when $s=0$ we have $y=y_0$ and
$z=z_0$, and thus $\rho ^2=y_0^2 + z_0^2=\rho _0^2$.
We easily see that (\ref{2.6}) satisfy (\ref{2.3}) and (\ref{2.4}) provided
\begin{eqnarray}\label{2.7}
X_1 &=&-\frac{1}{C} \left\{Y_1 y_1 + Z_1 z_1 +
\frac{1}{2}(Y_1^2 + Z_1^2)\right\} \ ,\nonumber\\
Y_1 &=& \frac{2 e e'}{m\rho_0 ^2}\, y_0\, ,\nonumber\\
Z_1 &=& \frac{2 e e'}{m\rho_0 ^2}\, z_0\, ,\nonumber\\
C&=& x_1 + t_1\, ,\nonumber\\
-1&=& x_1^2 + y_1^2 +z_1^2 -t_1^2\, .\nonumber\\
\end{eqnarray}
To verify that the first and fourth equation in (\ref{2.3}) are
satisfied we make use of the distributionally valid relation
$\theta (s)\, \delta (s)=\frac{1}{2}\, \delta (s)$,
for a justification of this relation in presence of an
impulsive gravitational wave see \cite{Stein}.

The relationship between $(\bar x,\bar y, \bar z, \bar
t)$ in $\bar {\cal S}$ and $(x, y, z, t)$ in ${\cal S}$ is given
by the Lorentz transformation
\begin{equation}\label{2.8}
\bar x=\gamma \,(x+v\,t)\ ,\qquad \bar y=y\
,\qquad \bar z=z\ ,\qquad \bar t=\gamma
\,(t+v\,x)\ ,
\end{equation}
with $\gamma =(1-v^2)^{-1/2}$. For the question currently under
consideration \emph{$v$ is close to unity} (the speed of light)
so that to the observer in ${\cal S}$ the field of the charge $e$
is given approximately by the impulsive wave (\ref{2.1}). To calculate
the deflection of the charge $e'$ in $\bar {\cal S}$ we take for
$s<0$ in ${\cal S}$: $x=x_0=0\
, y=y_0=-\eta\,(\eta >0)\ , z=z_0=0\ , t=s$ and so in view of
(\ref{2.6}) $\dot x=x_1=0\ , \dot
y=y_1=0\ , \dot z=z_1=0\ ,\dot t=t_1=1$. Then in $\bar {\cal
S}$ for $s<0$ we have, by (\ref{2.8}), $\bar x=\gamma v\,s\
, \bar y=-\eta\ , \bar z=0\ , \bar t=\gamma \,s$. It then
follows from (\ref{2.6}) that for $s>0$ we have in ${\cal S}$:
\begin{equation}\label{2.9}
x=X_1s\ ,\qquad y=-\eta +Y_1s\ ,\qquad z=Z_1s=0\ ,\qquad
t=(1-X_1)\,s\ .
\end{equation}
We note that with $z_0=0 , \rho _0=\eta$ and the third equation in (\ref{2.7})
yields $Z_1=0$. Thus using (\ref{2.8}) again we find that in $\bar {\cal S}$ for $s>0$:
\begin{eqnarray}\label{2.10}
\bar x&=&\gamma \,\{X_1+v\,(1-X_1)\}\,s\ ,\nonumber\\ \bar
y&=&-\eta+Y_1s\ ,\nonumber\\ \bar z&=&0\ ,\nonumber\\
\bar t&=&\gamma \,\{(1-X_1)+v\,X_1\}\,s\ .
\end{eqnarray}
The direction of motion of $e'$ in $\bar {\cal S}$ is the angle
the 3--velocity vector of $e'$ makes with the $\bar x$--direction.
If this angle is $\alpha$ (say) then for $s<0\ , \tan\alpha
=\dot{\bar y}/\dot{\bar x}=0$ while for $s>0$ we see from
(\ref{2.10}) that
\begin{equation}\label{2.11}
\tan\alpha =\frac{Y_1}{\gamma
\,\{v+(1-v)\,X_1\}}\ .
\end{equation}
with in our case $v$ close to unity. By (\ref{2.7}) with the conditions for
$s<0$ that we have chosen we have $X_1=-\frac{1}{2}Y_1^2$ and $Y_1=-\frac{2e\,e'}{m\,\eta}$.
Hence for $v$ close to unity (\ref{2.11}) gives the small angle of deflection
$\alpha$ of the high speed charge $e$ to be
\begin{equation}\label{2.12}
\alpha =-\frac{2\,ee'}{m\gamma \,\eta}\ ,
\end{equation}
This is positive if $ee'<0$. To check that we have arrived here at the correct angle of
scattering for a high speed charge $e$ scattered by a nucleus $e'$ we can, for example,
consult Synge \cite{S}. Synge's
exact formula for the angle of scattering is
\begin{equation}\label{2.13}
\alpha =-\pi +\frac{4}{K}\,\phi _1\ ,
\end{equation}
where
\begin{equation}\label{2.14}
K=\sqrt{1-\frac{k^2}{\eta ^2(\gamma ^2-1)}}\ ,
\end{equation}
with $k=-ee'/m$ and $\phi _1$ given by
\begin{equation}\label{2.15}
\tan ^2\phi _1=\frac{\sqrt{k^2+\eta ^2(\gamma ^2-1)^2}+k\,\gamma
_0}{\sqrt{k^2+\eta ^2(\gamma ^2-1)^2}-k\,\gamma }\ .
\end{equation}
For large $\gamma $ this gives approximately
\begin{equation}\label{2.16}
K=1\ ,\qquad \tan\phi _1=1+\frac{k}{\eta\,\gamma }\ ,
\end{equation}
and from (\ref{2.13}) we have
\begin{equation}\label{2.17}
\alpha =\frac{2\,k}{\eta\,\gamma }\ ,
\end{equation}
which agrees with (\ref{2.12}).

\setcounter{equation}{0}
\section{Preliminary Results for Gravitational Examples}\indent
For the remainder of this paper we will consider $\bar{\cal S}$ to
be the rest--frame of a gravitating source and ${\cal S}$ to be
the rest--frame of a high speed test particle. By analogy with the
electromagnetic example above we shall calculate the deflection of
a highly relativistic test particle moving in a gravitational
field. In section 4 this deflection is evaluated explicitly for a
high speed test particle moving in the Schwarzschild field, the
Kerr field, the field of a static, axially symmetric multipole
source and the field of a cosmic string. To the high speed
observer in the frame ${\cal S}$ all of these examples are
gravitational fields which resemble the gravitational fields of
different plane impulsive gravitational waves
\cite{BH}--\cite{BH2}. The line--element of the space--time model
in all cases takes the Kerr--Schild form
\begin{equation}\label{3.1}
ds^2=dx^2+dy^2+dz^2-dt^2-2H\,(dx+dt)^2\ ,
\end{equation}
with
\begin{equation}\label{3.2}
H=h(y\ , z)\,\delta (x+t)\ ,\end{equation} and the function $h(y
,z)$ differs from example to example (see section 4 below).

The time--like geodesic world--line $x^\mu =x^\mu (s)$ of a test particle in the space--time
with line--element given by (\ref{3.1}) and (\ref{3.2}) is given by the geodesic equations
\begin{eqnarray}\label{3.3}
\ddot x&=&H_x(\dot x+\dot t)^2+2\,(h_y\,\dot y+h_z\,\dot z)\,(\dot x+\dot t)\,\delta (x+t)\ ,
\nonumber \\
\ddot y&=&-h_y\,(\dot x+\dot t)^2\delta (x+t)\ ,\nonumber\\
\ddot z&=&-h_z\,(\dot x+\dot t)^2\delta (x+t)\ ,\nonumber\\
\ddot t&=&-H_x(\dot x+\dot t)^2-2\,(h_y\,\dot y+h_z\,\dot z)\,(\dot x+\dot t)\,\delta (x+t)\ .
\nonumber \\
\end{eqnarray}
Here as before the dot indicates differentiation with respect to
proper--time $s$. The subscripts on $H$ and $h$ indicate partial
derivatives with, in particular, $H_x=h(y ,z)\,\delta '(x+t)$ with
the prime denoting differentiation of the delta function with
respect to its argument. Particle motions in pp--wave space--times
(including impulsive waves and spinning particles) have been
studied in \cite{GV}, \cite{M}, \cite{T}, \cite{G} for example,
but none of these authors has considered the deflection question
that we are examining here. Following from (\ref{3.3}) we have the
first integral
\begin{equation}\label{3.4}
(\dot x+\dot t)\,\{\dot x-\dot t-2\,H\,(\dot x+\dot t)\}+\dot y^2+\dot z^2=-1\ .
\end{equation}
Adding the first and last equations in (\ref{3.3}) leads to $\ddot x
+\ddot t=0$ and so we have
\begin{equation}\label{3.5}
x+t=C\,s\ ,
\end{equation}
where $C$ is a constant of integration and we choose the point of intersection
of the world--line of the test particle with the null hyperplane $x+t=0$
to correspond to $s=0$. We also take $s<0$ to the past of $x+t=0$ and
$s>0$ to the future of $x+t=0$. We can now substitute into (\ref{3.3})
$\delta (x+t)=C^{-1}\delta (s)$ and $\delta '(x+t)=C^{-2}\delta '(s)$. It
follows from (\ref{3.3}) that for $s>0$ and for $s<0$ the time--like
world--line we are seeking is a geodesic of Minkowskian space--time with line--
element given by (\ref{3.1}) with $H=0$. As a result and in view of the form
of the right hand sides of (\ref{3.3}) we look for solutions of (\ref{3.3})
and (\ref{3.4}) which take the form
\begin{eqnarray}\label{3.6}
x&=&x_0+x_1s+X_1s\,\theta (s)+\hat X_1\theta (s)\ ,\nonumber\\
y&=&y_0+y_1s+Y_1s\,\theta (s)\ ,\nonumber\\
z&=&z_0+z_1s+Z_1s\,\theta (s)\ ,\nonumber\\
t&=&-x_0+t_1s-X_1s\,\theta (s)-\hat X_1\theta (s)\ ,\nonumber\\
\end{eqnarray}
with the coefficients of $s^0 , s , s\,\theta (s)$ and $\theta (s)$ here
all constants. We note that when $s=0$ we have $y=y_0\ , z=z_0$ and of
course $x+t=0$. In addition (\ref{3.5}) and (\ref{3.6}) imply that
\begin{equation}\label{3.7}
C=x_1+t_1\ .
\end{equation}
Substituting (\ref{3.6}) into (\ref{3.4}) gives
\begin{eqnarray}\label{3.8}
\hat X_1&=&h(y_0 , z_0)\equiv (h)_0\ ,\nonumber\\
x_1^2&+&y_1^2+z_1^2-t_1^2=-1\ ,\nonumber\\
2\,C\,X_1&+&2\,y_1Y_1+2\,z_1Z_1+Y_1^2+Z_1^2=0\ .\nonumber\\
\end{eqnarray}
Now the first of (\ref{3.3}) and also the fourth of (\ref{3.3})
are satisfied (verification of this involves using $\theta
(s)\,\delta (s) =\frac{1}{2}\delta (s)$ and $f(s)\,\delta
'(s)=-f'(0)\,\delta (s)$) while the second and third equations in
(\ref{3.3}) yield
\begin{equation}\label{3.9}
Y_1=-C\,\left (h_y\right )_0\ ,\qquad Z_1=-C\,\left (h_z\right )_0\ .
\end{equation}
As in the first of (\ref{3.8}) the brackets around a quantity followed
by a subscript zero denote that the quantity is evaluated at $y=y_0 ,
z=z_0$ which corresponds to $s=0$. By (\ref{3.9}) we can rewrite the
last of (\ref{3.8}) as
\begin{equation}\label{3.10}
X_1=\left (h_y\right )_0y_1+\left (h_z\right )_0z_1-\frac{C}{2}\,\left \{
\left (h_y\right )^2_0+\left (h_z\right )^2_0\right \}\ .
\end{equation}

A striking difference between (\ref{3.6}) and (\ref{2.6}) is the
appearance of the discontinuous $\hat X_1$--term in (\ref{3.6}). It
is essential to have this term present in the gravitational case. To see this
we first note from (\ref{3.6}) the simple limits
\begin{eqnarray}\label{3.11}
\lim_{s\rightarrow 0^-}x&=&x_0\ ,\qquad \lim_{s\rightarrow 0^-}t=-x_0\ ,
\nonumber\\
\lim_{s\rightarrow 0^+}x&=&x_0+\hat X_1\ ,\qquad \lim_{s\rightarrow 0^+}t
=-x_0-\hat X_1\ .\nonumber\\
\end{eqnarray}
From these we have
\begin{equation}\label{3.12}
\lim_{s\rightarrow 0^+}(x+t)=0=\lim_{s\rightarrow 0^-}(x+t)\ ,
\end{equation}
and
\begin{equation}\label{3.13}
\lim_{s\rightarrow 0^+}\left (\frac{x-t}{2}\right )=x_0+\hat X_1\
,\qquad \lim_{s\rightarrow 0^-}\left (\frac{x-t}{2}\right )=x_0\
.\end{equation} Now $v_+=\lim_{s\rightarrow 0^+}\left
(\frac{x-t}{2}\right )$ and $v_-=\lim_{s\rightarrow 0^-}\left
(\frac{x-t}{2}\right )$ are affine parameters along the generators
of $x+t=0\ (\Leftrightarrow s=0)$ on the future (plus) side and on
the past (minus) side respectively (see \cite{BH}, \cite{BH2}) and
by (\ref{3.13}) they are related by
\begin{equation}\label{3.14}
v_+=v_-+\hat X_1=v_-+(h)_0\ .
\end{equation}
This is precisely the matching condition required of these affine parameters
on the null hyperplane $x+t=0$ for this null hyperplane to be the history of
a gravitational wave. For this reason the presence of the discontinuous
$\hat X_1$--terms in (\ref{3.6}) is not surprising.

To set up the deflection problem for a high speed test particle
projected with 3--velocity $v$ in the positive $\bar x$--direction
for detailed consideration in section 4 we take for $s<0$ in
${\cal S}$: $x_0=0\ , x_1=0\ , y_1=0\ , z_1=0\ , t_1=1$ and thus
by (\ref{3.7}) $C=1$. Hence (\ref{3.6}) gives $x=0\ , y=y_0\ ,
z=z_0\ , t=s$. Using the Lorentz transformation (\ref{2.8}), with
as always $v$ assumed close to unity, we have in $\bar{\cal S}$
for $s<0$
\begin{equation}\label{3.15}
\bar x=\gamma\,v\,s\ ,\qquad \bar y=y_0\ ,\qquad \bar z=z_0\ ,\qquad
 \bar t=\gamma\,s\ .
\end{equation}
Now for $s>0$ in ${\cal S}$ we have
\begin{equation}\label{3.16}
x=X_1s+\hat X_1\ ,\qquad y=y_0+Y_1s\ ,\qquad z=z_0+Z_1s\ ,\qquad
t=(1-X_1)\,s-\hat X_1\ ,
\end{equation}
while in $\bar{\cal S}$ for $s>0$
\begin{eqnarray}\label{3.17}
\bar x&=&\gamma\,\left\{(1-v)\hat X_1+[v+(1-v)\,X_1]\,s\right\}\ ,
\nonumber\\
\bar y&=&y_0+Y_1s\ ,\nonumber\\
\bar z&=&z_0+Z_1s\ ,\nonumber\\
\bar t&=&\gamma\,\left\{(v-1)\hat X_1+[1-(1-v)\,X_1]\,s\right\}\ .
\nonumber\\
\end{eqnarray}
If $\alpha$ is the angle of deflection of the high speed test particle
out of the $\bar x\,\bar z$--plane for $s>0$ and if $\beta$ is the angle of
deflection out of the $\bar x\,\bar y$--plane for $s>0$ then these angles
are given by
\begin{equation}\label{3.18}
\tan\alpha =\frac{\dot{\bar y}}{\sqrt{\dot{\bar x}^2+\dot{\bar z}^2}}=
\frac{Y_1}{\left \{Z_1^2+\gamma ^2[X_1\,(1-v)+v]^2\right\}^{1/2}}\ ,
\end{equation}
and
\begin{equation}\label{3.19}
\tan\beta =\frac{\dot{\bar z}}{\dot{\bar x}}=\frac{Z_1}{\gamma\,[X_1\,(1-v)+v]}
\ ,
\end{equation}
where now (\ref{3.9}) and (\ref{3.10}) reduce to
\begin{eqnarray}\label{3.20}
X_1&=&-\frac{1}{2}\left [\left (h_y\right )^2_0+\left (h_z\right )^2_0\right ]\ ,
\nonumber\\
Y_1&=&-\left (h_y\right )_0\ ,\qquad Z_1=-\left (h_z\right )_0\ .
\nonumber\\
\end{eqnarray}
The equations (\ref{3.18}) and (\ref{3.19}) hold for $v$ close to
unity. For the explicit examples of the function $h(y, z)$ in
section 4 we will be able to evaluate (\ref{3.18}) and
(\ref{3.19}) for $v$ close to unity and thereby obtain the
dominant terms in (\ref{3.18}) and (\ref{3.19}).

\setcounter{equation}{0}
\section{Gravitational Examples}\indent
We now specialise the formulas (\ref{3.18}) and (\ref{3.19}) giving the
deflection of a highly relativistic test particle to some specific
gravitational fields. We begin with the Kerr gravitational field. When this
field is boosted to the speed of light in the negative $x$--direction the
line--element of the space--time is given by (\ref{3.1})
and (\ref{3.2}) with (see \cite{BH2})
\begin{equation}\label{4.1}
h=2\,p\,\log\{(y-a)^2+z^2\}\ .
\end{equation}
Here $a$ is the angular momentum per unit mass of the Kerr source and
$p=m\,\gamma$ with $m$ the rest--mass of the Kerr source. The non--identically
vanishing components of the Riemann tensor in this case are proportional
to $\delta (x+t)$ and their coefficients are singular at $y=a\ , z=0$ on
the null hyperplane $x+t=0$ which is the history of an impulsive
gravitational wave. We note that $y=a\ , z=0$ is a null geodesic generator
of $x+t=0$.

When (\ref{4.1}) is
substituted into (\ref{3.18}) and (\ref{3.19}) we obtain the dominant
terms by taking $v$ close to unity.
In this case the deflection angles $\alpha$ and $\beta$ are
given by
\begin{equation}\label{4.2}
\tan\alpha =\frac{-4m(y_0-a)}{\left\{\left [(y_0-a)^2+z_0^2-4m^2\right ]^2+16m^2z_0^2\right\}^{1/2}}\ ,
\end{equation}
and
\begin{equation}\label{4.3}
\tan\beta =\frac{-4mz_0}{(y_0-a)^2+z_0^2-4m^2}\ .
\end{equation}
If a particle is projected from a point in the equatorial plane ($z_0=0$)
then (\ref{4.3}) shows that $\beta = 0$ and thus the particle will
remain in the equatorial plane.
We see from (\ref{4.2}) that $\alpha =\pi /2$ corresponds to $y_0 =\pm 2m+a$.
This is the value of $y_0$ corresponding to the capture of the incoming
particle in this case.
If such a particle is projected
from a point on the $y$--axis at which $y_0=-\eta \ (\eta >0)$ then
the angle
of deflection $\alpha$ is given by
\begin{equation}\label{4.4}
\tan\alpha =\frac{4m(\eta +a)}{\left [(\eta +a)^2-4m^2\right ]}\ ,
\end{equation}
and for large values of $\eta$ this gives the small deflection angle
\begin{equation}\label{4.5}
\alpha =\frac{4m}{\eta}-\frac{4ma}{\eta ^2}\ .
\end{equation}
This formula agrees with the small angle of deflection of a photon
path in the equatorial plane of the Kerr source calculated by
Boyer and Lindquist \cite{BL} (in \cite{BL} the angular momentum
of the source points in the negative $z$--direction). We note that
the leading term in (\ref{4.5}) is Einstein's small angle $\alpha
_0$ of deflection of light in the Schwarzschild field. The effect
of the second term in (\ref{4.5}) is to \emph{decrease} the
deflection angle from $\alpha _0$ if $a>0$
 and
to \emph{increase} the deflection angle from $\alpha _0$ if $a<0$.
For the unboosted Kerr source, $a>0$ here corresponds to angular
momentum in the positive $z$--direction (see \cite{BH2} and
\cite{DKS}).

\begin{figure}
\center
\includegraphics[width=8.5cm]{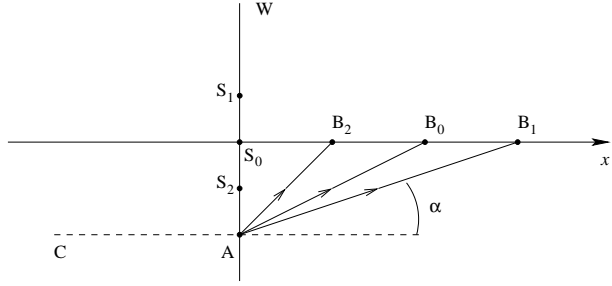}
\caption{\label{1} $CA$ is the direction of the incoming highly relativistic
test particle. $W$ is
the plane impulsive gravitational wave front travelling with the
speed of light in the negative $x$--direction. $AB_0$ is the deflected
path of the particle when $a=0$ and the singularity in the wave front
is at $S_0$. $AB_1$ is the deflected path when $a>0$ and $S_1$ is the
singularity in the wave front. $AB_2$ is the deflected path when $a<0$
and $S_2$ is the singularity in the wave front.}
\end{figure}

It follows from figure 1 that $a>0$ means that the singularity
$y=a\ , z=0$ in the impulsive gravitational wave front is further
from the incoming particle than it would be in the Schwarzschild
case $(a=0)$ and thus the deflection is decreased from its
Schwarzschild value. In this case from the point of view of the
original Kerr gravitational field the incoming particle passes the
black hole in the same anti--clockwise direction as the hole's
rotation. If $a<0$ the singularity is closer to the incoming
particle and the deflection is increased. Now the particle passes
the black hole in the opposite direction to its (anti--clockwise)
rotation.

\begin{figure}
\center
\includegraphics[width=8.5cm]{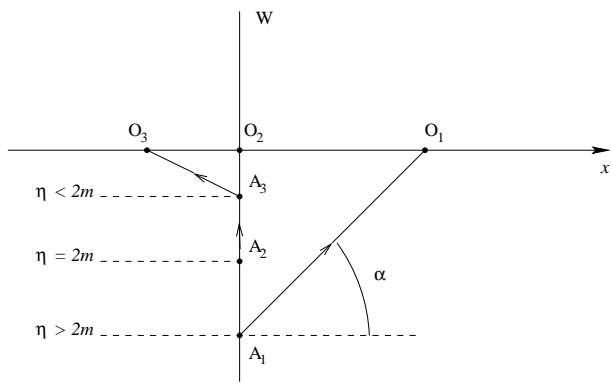}
\caption{\label{2} The dotted lines are the directions of highly relativistic
incoming particles corresponding to the values of $\eta$ indicated.
$A_1 O_1 , A_2 O_2$ and $A_3 O_3$ are the corresponding deflected paths
for which the values of the deflection angle are less than, equal to
and greater than $\pi /2$ respectively.}
\end{figure}

The situation corresponding to deflection in the
Schwarzschild field $(a=0)$ is described in figure 2.
In this figure $W$
is the plane impulsive gravitational wave front and the singular point
 $(y=a\ , z=0)$ is $O_2$. This figure indicates how the capture of a high
speed particle by a Schwarzschild black hole manifests itself in
the current context. Our calculations are most accurate for large
values of $\eta$. Nevertheless we see from the formula (\ref{4.4})
with $a=0$ that the analogue here of the capture cross-section
radius is $\eta =2\,m$ ($\Leftrightarrow\alpha =\pi /2$). This is
smaller than the $3\sqrt{3}\,m$ capture cross--section radius for
light in the Schwarzschild field because the highly relativistic
particle does not have to pass through a gravitational field
before being scattered (see \cite{VF} where the capture
cross--section radius is given for Schwarzschild (p.48) and for
Kerr (p.77)).

To consider from our point of view the deflection of a highly
relativistic particle in the field of a static, axially symmetric,
multipole source we need the light--like boosted form of this
gravitational field. This form is given by the line--element
(\ref{3.1}) and (\ref{3.2}) with (see \cite{BH})
\begin{equation}\label{4.6}
h=2\,\sum_{l=0}^{\infty}p_l\,\frac{(-1)^l}{l!}\,\frac{\partial
^lg} {\partial z^l}\ ,
\end{equation}
with $g=\log (y^2+z^2)$ and $p_l=\gamma\,A_l\ (l=0, 1, 2, \dots )$
where $A_l$ are simply related to the multipole moments of the
source (with $A_0=m$ its rest--mass). The deflection angles given
by (\ref{3.18}) and (\ref{3.19}) can now be calculated and $v$
taken close to unity in these expressions. If for simplicity we
project the high speed particle from a point in the plane $z_0=0$
lying on the $y$--axis at which $y_0=-\eta \ (\eta
>0)$ then the formulas (\ref{3.18}) and (\ref{3.19}) yield for $v$
close to unity
\begin{eqnarray}\label{4.7}
\tan\alpha &=&\frac{\tilde Y_1}{\left \{(1+\frac{1}{2}\tilde X_1)^2+\tilde Z_1^2\right\}^{1/2}}\ ,\nonumber\\
\tan\beta &=&\frac{\tilde Z_1}{1+\frac{1}{2}\tilde X_1}\ ,\nonumber\\
\end{eqnarray}
with
\begin{equation}\label{4.8}
\tilde Y_1=4\sum_{l=0}^{\infty}A_{2l}\,\frac{(-1)^l}{\eta ^{2l+1}}\ ,
\qquad \tilde Z_1=-4\sum_{l=0}^{\infty}A_{2l+1}\,\frac{(-1)^{l+1}}
{\eta ^{2l+2}}\ '
\end{equation}
and
\begin{equation}\label{4.9}
\tilde X_1=-\frac{1}{2}\left (\tilde Y_1^2+\tilde Z_1^2\right )\ .
\end{equation}
We see that $\tilde Z_1\neq 0$ involves the \emph{odd}--pole
moments of the original source of the gravitational field and is
responsible for the deflection of a highly relativistic particle
\emph{out} of the plane $z=0$. Also $\tilde Y_1\neq 0$ involves
only the \emph{even}--pole moments of the source and is
responsible for deflection \emph{in} the plane $z=0$.

Finally the light--like boosted gravitational field of a cosmic string
\cite{BHI} is given by the line--element (\ref{3.1}) and (\ref{3.2})
if $h(y, z)$ is taken to be
\begin{equation}\label{4.10}
h=4\pi\,\tilde \mu\,|y|\ ,
\end{equation}
and $\tilde \mu =\gamma\,\mu$ with $\mu$ the mass per unit length of
the original string. This results in the angle $\beta =0$ as expected
because of the infinite line source along the $z$--axis. After taking
$v$ close to unity we find that
\begin{equation}\label{4.11}
\tan\alpha =\frac{4\pi\,\mu\,\left (2\theta (y_0)-1\right )}
{4\pi ^2\mu ^2-1}\ .
\end{equation}
At first sight it would appear from this that irrespective of the choice of
$y_0$ we could have $\alpha =\pi /2$ simply by taking $\mu =(2\,\pi )^{-1}$.
However this is not possible because the deficit angle $\Delta\phi$
associated with the cosmic string satisfies
\begin{equation}\label{4.12}
\Delta\phi =8\pi\,\mu <\pi\ .
\end{equation}

\setcounter{equation}{0}
\section{A High Speed Test Particle with Spin}\indent
\setcounter{equation}{0} We now generalise the results of the
previous two sections by allowing the high speed test particles to
have intrinsic spin. Writing the 4--velocity of the particle as
$\dot x^\alpha =dx^\alpha /ds$, its 4--momentum as $p^\alpha$ and
its spin tensor as $S^{\alpha\beta}=-S^{\beta\alpha}$, these
quantities satisfy the equations of motion (see \cite{Di}
eqs.(7.3)--(7.5))
\begin{eqnarray}\label{5.1}
\frac{Dp^\alpha}{Ds}&=&-\frac{1}{2}\,R_{\delta\gamma\beta}{}^{\alpha}\,
S^{\delta\gamma}\,\dot x^\beta\ ,\\
\frac{DS^{\alpha\beta}}{Ds}&=&p^\alpha\,\dot x^\beta
-p^{\beta}\,\dot x^\alpha\ ,\\ p_{\alpha}\,S^{\alpha\beta}&=&0\ .
\end{eqnarray}
Here $D/Ds$ denotes covariant differentiation in the direction of
$\dot x^\alpha$ and $R_{\delta\gamma\beta\gamma}$ are the
components of the Riemann curvature tensor of the space--time
containing the history of the particle. We find it convenient to
work with the spin vector $S^\alpha$ associated with the particle,
which is equivalent to the spin tensor $S^{\alpha\beta}$ and is
defined by
\begin{equation}\label{5.4}
S^\alpha =\frac{1}{2}\,\eta
^{\alpha\beta\gamma\delta}\,p_{\beta}\,S_{\gamma\delta}\ ,
\end{equation}
where $\eta _{\alpha\beta\gamma\delta}=\sqrt{-g}\,\epsilon
_{\alpha\beta\gamma\delta}$ with $g={\rm det}(g_{\mu\nu})$ and
$\epsilon _{\alpha\beta\gamma\delta}$ is the 4--dimensional
Levi--Civita permutation symbol. In terms of the spin vector
(5.1)--(5.3) can be written
\begin{eqnarray}\label{5.5}
\frac{Dp^\alpha}{Ds}&=&-(p_{\sigma}\,p^{\sigma})^{-1}\,{}^*
R^{\mu\nu}{}_{\beta}{}^\alpha\,p_\mu\,S_\nu\,\dot x^\beta\ ,\\
\frac{DS^\alpha}{Ds}&=&(p_{\sigma}\,p^{\sigma})^{-2}\,{}^*R^{\mu\nu}
{}_{\rho\beta}\,p_{\mu}\,S_{\nu}\,\dot x^\rho\,(p^\alpha\,S^\beta
-
p^\beta\,S^\alpha )\ ,\\ p_\alpha\,S^\alpha &=&0\ .
\end{eqnarray}
Here ${}^*R^{\mu\nu}{}_{\rho\beta}=\frac{1}{2}\,\eta
^{\mu\nu\lambda\sigma}\,R_{\lambda\sigma\rho\beta}$ is the left
dual of the Riemann tensor. In general the 4--momentum $p^\alpha$
is not collinear with the 4--velocity $\dot x^\alpha$. However if
we consider a \emph{slowly} spinning particle, in the sense that
we neglect squares and higher powers of the spin vector
components, then we can write \cite{Di}
\begin{equation}\label{5.8}
p^\alpha =m_0\,\dot x^\alpha\ ,
\end{equation}
where $m_0$ is the constant rest--mass of the particle. We will
exhibit here a solution of (5.5)--(5.7), in the space--time with
line--element given by (3.1) and (3.2), for the special case in 
which (\ref{5.8}) holds exactly provided the spin vector takes
a special form given below. 
In this case (5.5)--(5.7), written out
explicitly, simplify to
\begin{eqnarray}\label{5.9}
\ddot x^\alpha +\Gamma ^\alpha _{\beta\sigma}\,\dot x^\beta\,\dot
x^\sigma &=&-\frac{1}{m_0}\,{}^*R^{\alpha}{}_{\beta\mu\nu}\,\dot
x^\beta\,\dot x^\mu\,S^\nu\ ,\\ \dot S^\alpha +\Gamma ^\alpha
_{\beta\sigma}\,S^\beta\,\dot x^\sigma
&=&\frac{1}{m_0^2}\,{}^*R_{\mu\nu\rho\beta}\,\dot
x^\mu\,S^\nu\,\dot x^\rho\,S^\beta\,\dot x^\alpha\ ,\\ \dot
x^\alpha\,S_\alpha &=&0\ ,
\end{eqnarray}
where, as in the previous sections, a dot denotes differentiation
with respect to proper--time $s$. $\Gamma ^\alpha _{\beta\sigma}$
are the components of the Riemannian connection. The equations
(5.9) have the first integral
\begin{equation}\label{5.12}
g_{\alpha\beta}\,\dot x^\alpha\,\dot x^\beta =-1\ .
\end{equation}We apply equations (5.9)--(5.11) to the motion of a
high speed, spinning test particle moving in the equatorial plane
$z=0$ of the Schwarzschild and Kerr sources (in which case $H_z$
vanishes on $z=0$) for which $S^\alpha =m_0\,S\,\delta
^{\alpha}_{3}$ with $S$ a constant. Here $S$ is the angular
momentum per unit mass of the spinning particle and it is
collinear with the angular momentum of the black--hole in the Kerr
case (for other examples of spinning test particle motion in the
Kerr gravitational field in which the particle spin is collinear
with the spin of the black--hole see \cite{RW} and \cite{Tod}; for
more general motions of spinning test particles in the Kerr field
see \cite{Sem}). For the line--element given by (\ref{3.1}) and
(\ref{3.2}) the right hand side of (5.10) now vanishes and thus
the spin propagation equations, written out in full, read
\begin{eqnarray}\label{5.13}
&&\dot S^1-\left\{H_x\,(\dot x+\dot t)+\dot y\,H_y+\dot
z\,H_z\right\}\,(S^1+S^4)\nonumber\\&&-H_y\,(\dot x+\dot
t)\,S^2-H_z\,(\dot x+\dot t)\,S^3=0\ ,\end{eqnarray}
\begin{equation}\label{5.14} \dot S^2+H_y\,(\dot x+\dot
t)\,(S^1+S^4)=0\ ,\end{equation}
\begin{equation}\label{5.15} \dot S^3+H_z\,(\dot x+\dot t)\,(S^1+S^4)=0\ ,
\end{equation}
\begin{eqnarray}\label{5.16}
&&\dot S^4+\{H_x\,(\dot x+\dot t)-\dot y\,H_y-\dot
z\,H_z\}\,S^1-H_y\,(\dot x-\dot t)\,S^2\nonumber\\&&-H_z\,(\dot
x-\dot t)\,S^3+\{H_x\,(\dot x+\dot t)+\dot y\,H_y+\dot
t\,H_z\}\,S^4=0\ .\end{eqnarray} The subscripts on $H$ indicate
partial derivatives as before and on account of (\ref{3.2}) we
have $H_x=H_t$. It is easy to see that these equations are
satisfied on $z=0$ with $S^\alpha =m_0\,S\,\delta ^{\alpha}_{3}$,
$S$ a constant and $H_z=0$. With $S^\alpha =m_0\,S\,\delta
^{\alpha}_{3}$ the equations (\ref{5.9}), calculated using the
metric tensor given via (\ref{3.1}) and (\ref{3.2}), read
\begin{equation}\label{5.17}
\ddot x-H_x\,(\dot x+\dot t)^2-2\,(\dot y\,H_y+\dot z\,H_z)\,(\dot
x+\dot t)+S\,(\dot y\,H_{yy}+\dot z\,H_{yz})\,(\dot x+\dot t)=0\ ,
\end{equation}
\begin{equation}\label{5.18}
\ddot y+H_y\,(\dot x+\dot t)^2-S\,H_{yy}\,(\dot x+\dot t)^2=0\ ,
\end{equation}
\begin{equation}\label{5.19}
\ddot z+H_z\,(\dot x+\dot t)^2-S\,H_{yz}\,(\dot x+\dot t)^2=0\ ,
\end{equation}
\begin{equation}\label{5.20}
\ddot t+H_x\,(\dot x+\dot t)^2+2\,(\dot y\,H_y+\dot z\,H_z)\,(\dot
x+\dot t)-S\,(\dot y\,H_{yy}+\dot z\,H_{zz})\,(\dot x+\dot t)=0\ .
\end{equation}
These equations specialise to (\ref{3.3}) when $S=0$ and the
technique for solving them is the same as that employed in section
3. The result is that the equations (\ref{3.20}), for the case we
are considering in which $(H_z)_0=0=(h_z)_0$, now read
\begin{eqnarray}\label{5.21}
X_1&=&-\frac{1}{2}\left [(h_y)^2_0-2\,S\,(h_y)_0\,(h_{yy})_0\right
]\ ,\nonumber\\ Y_1&=&-(h_y)_0+S\,(h_{yy})_0\ ,\nonumber\\
 Z_1&=&0\ .
\end{eqnarray}
It is easy to see that (5.11) is now automatically satisfied.
Using $h(y, z)$ for the Kerr example given by (\ref{4.1}), the
angle of deflection of the high speed, spinning test particle
moving in the equatorial plane is given by the following
generalisation of (\ref{4.4}):
\begin{equation}\label{5.22}
\tan\alpha =\left [1-\frac{4\,m^2}{(\eta
+a)^2}+\frac{4\,m^2\,S}{(\eta +a)^3}\right ]^{-1}\,\left (
\frac{4\,m}{\eta +a}-\frac{4\,m\,S}{(\eta +a)^2}\right )\ .
\end{equation}
For large values of the impact parameter $\eta$ we thus have the
small deflection angle
\begin{equation}\label{5.23}
\alpha =\frac{4\,m}{\eta}-\frac{4\,m\,(a+S)}{\eta ^2}\ .
\end{equation}
A notable feature of this formula is that if the spin of the test
particle is equal and opposite to the spin of the black--hole then
the test particle is deflected through the same angle as that of
light in the Schwarzschild field (cf. (\ref{4.5}) with $a=0$). In
the case of a Weyl source with multipole moments, $z=0$ is the
equatorial plane provided the odd--pole moments vanish. In this
case the function $h(y, z)$ appearing in (5.21) is given by (4.6)
with $p_{2l+1}=0$, $l=0, 1, 2, \dots\ \ $. For large impact
parameter the angle of deflection of the high speed spinning test
particle in this case is given by (\ref{5.23}) with $a=0$.

\setcounter{equation}{0}
\section{Discussion}\indent
The calculations of the angles of deflection $\alpha$ and $\beta$
described in section 4 make use of the \emph{spatial} components
of the 4--velocity of the highly relativistic test particle before
and after it encounters the electromagnetic or gravitational field
which is responsible for the deflection. There is also interesting
information contained in the time components of the 4--velocity.
For example in the electromagnetic case in section 2 we see from
(\ref{2.9}) that in the frame ${\cal S}$ for $s>0$ we have $\dot
t=1-X_1$ while for $s<0$ we have $\dot t=1$. Hence the energy
acquired by the high speed charge $e'$ of rest--mass $m$ after
scattering off the nucleus of charge $e$ is
\begin{equation}\label{6.1}
\Delta E=-m\,X_1=\frac{1}{2}Y_1^2=\frac{2 (e'e)^2}{m\,\eta ^2}\ .
\end{equation}
This agrees with Eq.(13.2) in \cite{J} for the acquired energy
when the speed of the incoming particle is close to the speed of
light. If $m_0$ is the rest-mass of the high speed test particle
scattered by the Kerr gravitational field and considered in
section 4,then the corresponding formula to (\ref{6.1}) in this
case is easily seen to be
\begin{equation}\label{6.2}
\Delta E=\frac{8m_0m^2\gamma ^2}{(\eta +a)^2}\ ,
\end{equation}
for $v$ close to unity.

\end{document}